# Ba$_2$NiOsO$_6$: a Dirac-Mott insulator with ferromagnetism near 100 K

Hai L. Feng,[a,b,*] Stuart Calder,[c] Madhav Prasad Ghimire,[d,e,*] Ya-Hua Yuan,[a,f] Yuichi Shirako,[g,#] Yoshihiro Tsujimoto,[a] Yoshitaka Matsushita,[h] Zhiwei Hu,[b] Chang-Yang Kuo,[b] Liu Hao Tjeng,[b] Tun-Wen Pi,[i] Yun-Liang Soo,[i,j] Jianfeng He,[a,f] Masahiko Tanaka,[k] Yoshio Katsuya,[k] Manuel Richter,[d,l] Kazunari Yamaura [a,f]

[a] Research Center for Functional Materials, National Institute for Materials Science, 1-1 Namiki, Tsukuba, Ibaraki 305-0044, Japan

[b] Max Planck Institute for Chemical Physics of Solids, Nöthnitzer Str. 40, 01187 Dresden, Germany

[c] Quantum Condensed Matter Division, Oak Ridge National Laboratory, Oak Ridge, Tennessee 37831, USA.

[d] Leibniz Institute for Solid State and Materials Research, IFW Dresden, PO box 270116, D-01171 Dresden, Germany

[e] Condensed Matter Physics Research Center, Butwal-13, Rupandehi, Lumbini, Nepal

[f] Graduate School of Chemical Sciences and Engineering, Hokkaido University, North 10 West 8, Kita-ku, Sapporo, Hokkaido 060-0810, Japan

[g] Department of Chemistry, Gakushuin University, 1-5-1 Mejiro, Toshima-ku, Tokyo 171-8588, Japan

[h] Materials Analysis Station, National Institute for Materials Science, 1-2-1 Sengen, Tsukuba, Ibaraki 305-0047, Japan

[i] National Synchrotron Radiation Research Center, Hsinchu 30076, Taiwan

[j] Department of Physics, National Tsing Hua University, Hsinchu 30013, Taiwan

[k] Synchrotron X-ray Station at SPring-8, National Institute for Materials Science, Kouto 1-1-1, Sayo-cho, Hyogo 679-5148, Japan

[l] Dresden Center for Computational Materials Science, DCMS, TU Dresden, D-01069 Dresden, Germany

[*] (Corresponding authors) E-mail: Hai.FENG_nims@hotmail.com (HF) and ghimire.mpg@gmail.com (MPG)

[#] (Present Addresses) Department of Crystalline Materials Science, Nagoya University, Furo-cho, Chikusa-ku, Nagoya, Aichi 464-8603, Japan

**ABSTRACT**

The ferromagnetic semiconductor $Ba_2NiOsO_6$ ($T_{mag}$ ~100 K) was synthesized at 6 GPa and 1500 °C. It crystallizes into a double perovskite structure [$Fm$-$3m$; $a = 8.0428(1)$ Å], where the $Ni^{2+}$ and $Os^{6+}$ ions are perfectly ordered at the perovskite B-site. We show that the spin-orbit coupling of $Os^{6+}$ plays an essential role in opening the charge gap. The magnetic state was investigated by density functional theory calculations and powder neutron diffraction. The latter revealed a collinear ferromagnetic order in a >21 kOe magnetic field at 5 K. The ferromagnetic gapped state is fundamentally different from that of known dilute magnetic semiconductors such as (Ga,Mn)As and (Cd,Mn)Te ($T_{mag} < 180$ K), the spin-gapless semiconductor $Mn_2CoAl$ ($T_{mag}$ ~720 K), and the ferromagnetic insulators EuO ($T_{mag}$ ~70 K) and $Bi_3Cr_3O_{11}$ ($T_{mag}$ ~220 K). It is also qualitatively different from known ferrimagnetic insulator/semiconductors, which are characterized by an antiparallel spin arrangement. Our finding of the ferromagnetic semiconductivity of $Ba_2NiOsO_6$ should increase interest in the platinum group oxides, because this new class of materials should be useful in the development of spintronic, quantum magnetic, and related devices.

# 1. INTRODUCTION

The study of dilute magnetic semiconductors has been motivated since the 1970s by the prospect that materials with tight coupling between magnetic and semiconducting properties may be useful in developing spintronic and related devices.[1-3] Early studies doped magnetic impurities into semiconducting hosts such as the III-V, II-VI, and IV semiconductors. Successful synthesis of a paramagnetic (II,Mn)-VI semiconductor was reported in 1978 followed by discovery of long range ferromagnetic (FM) order at ~10 K by Ohno *et al*. in 1992.[4] The magnetic ordering temperature ($T_{mag}$) increased to 100 K by 1998,[3] but the $T_{mag}$ of most dilute magnetic semiconductors to date remains at or below ~180 K.[5]

The variety of magnetic semiconductors has increased in concert with ongoing studies of dilute magnetic semiconductivity. For example, a "spin-gapless semiconductor" has been produced by synthesis of the Heusler alloy, $Mn_2CoAl$, which exhibits spin polarized transport at room temperature and holds promise for spintronic applications.[6] In 2007, a Co-doped $TiO_2$ film was claimed to show FM order at room temperature.[7] However, the essential mechanism of its ferromagnetism is still debated.[8, 9] Attempts to apply the popular FM insulator EuO ($T_{mag}$ ~70 K)[10] to technological use are in progress. $La_2NiMnO_6$ [11, 12] and $Bi_2NiMnO_6$ [13-15] have been studied, because they have the potential to act as a FM insulator in practical applications. $Bi_3Cr_{2.91}O_{11}$ recently has been suggested to be a FM insulator with $T_{mag}$ ~220 K.[16] In this study, we successfully synthesize an FM oxide, $Ba_2NiOsO_6$, in which the charge gap is produced by a mechanism fundamentally different from that of any proposed to date for known FM insulators/semiconductors such as above. The unconventional origin of this property suggests significant feasibility in designing a practical room temperature FM insulator because a few 3d-5d double perovskite oxides (DPOs) related to $Ba_2NiOsO_6$ also exhibit magnetic order far above room temperature (Table S1) [17] as well as related 4d and 5d oxides $SrTcO_3$ ($T_{mag}$ ~1000 K) [18] and $NaOsO_3$ ($T_{mag}$ = 410 K),[19] yet magnetic interactions are still poorly understood.

We report here the high-pressure, high-temperature synthesis of the cubic DPO, $Ba_2NiOsO_6$, and its characterization by measurement of bulk magnetic properties and by powder neutron and X-ray diffraction. A hexagonal phase of $Ba_2NiOsO_6$ ($P$-3$m$1; $a$ = 5.73 Å and $c$ = 14.07 Å) was synthesized without high pressure in 1980, but its magnetic properties were not investigated.[20] Density functional theory (DFT) calculations show that spin-orbit coupling (SOC) plays an essential role in generating the charge gap. To our knowledge, cubic $Ba_2NiOsO_6$ differs qualitatively from most ferrimagnetic (FIM) semiconductors, in which spins are ordered in an antiparallel fashion. Our findings point to a new class of FM insulators driven by a platinum group element that may be useful in the synthesis of practical magnetic semiconductors.

## 2. EXPERIMENTAL

**Materials synthesis:** Polycrystalline $Ba_2NiOsO_6$ was synthesized by solid state reaction at high pressure from powders of $BaO_2$ (99%, Kojundo Chemical Lab. Co., Ltd., Japan), NiO (99.99%, Kojundo Chemical Lab. Co., Ltd., Japan), Os (99.95%, Heraeus Materials Technology, Germany), and $KClO_4$ (>99.5%, Kishida Chemical Co., Ltd., Japan). The powders were thoroughly mixed in stoichiometric ratio ($BaO_2$/NiO/Os/$KClO_4$ = 2:1:1:0.25) and sealed in a Pt capsule within an Ar-filled glove box. The sealed capsule was gradually and isotropically compressed in a belt-type press (Kobe Steel, Ltd., Japan) to a pressure of 6 GPa. It was then heated at 1500 °C for 1 h, while maintaining the capsule pressure. After heating, the sample was quenched to 100 °C or below within several seconds followed by release of the pressure over a few hours. The resulting pellet was polycrystalline, dense, and well-sintered.

**X-ray and neutron diffraction analysis:** A piece of $Ba_2NiOsO_6$ pellet was finely ground and rinsed with water to remove residual KCl. The final fine black powder was investigated by synchrotron X-ray diffraction (SXRD) at the BL15XU beamline facility, SPring-8, Japan. SXRD data were collected at ambient temperature using a high precision Debye-Scherrer type diffractometer [21] and the "MYTHEN" one-dimensional detector.[22] The radiation wavelength was 0.65297 Å, which was confirmed by a $CeO_2$ reference. The SXRD pattern was analyzed by the Rietveld method using "RIETAN-VENUS" software.[23, 24] Neutron powder diffraction (ND) was performed on a 5-g sample of $Ba_2NiOsO_6$ powder using the HB-2A diffractometer at the High Flux Isotope Reactor at the Oak Ridge National Laboratory at temperatures down to 4 K. Two wavelengths were selected using a Ge monochromator. A wavelength of 1.54 Å was used to provide sufficient reflections for an accurate crystal structure determination. A 2.41 Å wavelength was used for the magnetic structure determination in both zero field and applied fields up to 45 kOe. The ND data were analyzed using the Rietveld refinement program "FullProf".[25] The magnetic structural representation analysis was performed using "SARAh".[26]

**Materials properties characterization:** X-ray absorption spectroscopy (XAS) at the Ni-$L_{2,3}$ and Os-$L_3$ edges was conducted at the BL07C and BL08B beamlines using the total electron yield and transmission method, respectively, in National Synchrotron Radiation Research Center (NSRRC) in Taiwan. The Ni-$L_{2,3}$ spectrum of NiO and the Os-$L_3$ spectrum of Sr$_2$FeOsO$_6$ [27, 28] were measured simultaneously as references for Ni$^{2+}$ and Os$^{5+}$, respectively.

Other pieces of the Ba$_2$NiOsO$_6$ pellet were subjected to electrical resistivity ($\rho$) and specific heat ($C_p$) measurements. The $\rho$ measurement was conducted at 100–300 K using a four-point method with a physical properties measurement system from Quantum Design, Inc. Electrical contacts were prepared longitudinally on a polycrystalline piece with Pt wires and silver paste. The DC gauge current was 0.1 mA. The temperature dependence of $C_p$ was measured at 2–300 K by a thermal relaxation method. The contribution from residual KCl to the raw $C_p$ data was subtracted using tabulated data[29] and the mole ratio (KCl:Ba$_2$NiOsO$_6$ = 0.25:1).

An amount of rinsed powder was subjected for magnetic studies; the magnetic susceptibility ($\chi$) of powder Ba$_2$NiOsO$_6$ was measured with a magnetic property measurement system from Quantum Design, Inc. conducted in field-cooled (FC) and zero-field-cooled (ZFC) modes between 2 and 390 K in fixed magnetic fields of 0.1, 10, and 50 kOe. Isothermal magnetization was measured at 5, 50, and 200 K in magnetic fields between -50 kOe and +50 kOe with the same apparatus. Note that the mass fraction of NiO was 2.4% according to the SXRD analysis (see Fig. 1), indicating a small impurity level. Besides, NiO is an established antiferromagnetic (AFM) insulator; therefore, an AFM contribution at such the low concentration to the bulk properties was expected to be trivial. We hence did not correct raw magnetic data of the magnetic susceptibility and magnetization measurements.

**Theoretical calculations:** DFT calculations were carried out using the full-potential linearized augmented plane wave method as implemented in the WIEN2k code.[30] The atomic sphere radii ($R_{MT}$) were fixed at 2.5, 2.1, 2.0, and 1.63 Bohr radii for Ba, Ni, Os, and O, respectively. The linear tetrahedron method with 3000 (1000, 500) $k$ points in the entire Brillouin zone was employed for the reciprocal space integrations in the case of a cubic unit cell containing one chemical unit cell (and two for tetragonal and four for orthorhombic unit cells). The standard generalized gradient approximation (GGA) in the parameterization of Perdew, Burke, and Ernzerhof (PBE-96) [31] was used in the initial calculations. The GGA+$U$ ($U$ = Coulomb interaction) functional with double counting corrections according to Anisimov *et al*. [32] (a variant of the so-called fully localized limit with $J = 0$) was applied for the main investigations. Values of $U = 5$ and 2 eV were used for the Ni and Os atoms, respectively. The calculations were tested by varying these values between 4 and 7 eV for Ni and 1 and 4 eV for Os, respectively. These ranges are comparable to values reported for 3d and 5d oxides (~2– 10 eV for 3d and ~1–3 eV for 5d).[33-35] SOC was considered via a second variational step using the scalar relativistic eigenfunctions as basis. All calculations were conducted using the experimental lattice parameters obtained at 4 K. Parts of the calculations were double checked with the Full Potential Local Orbital (FPLO) code.[36] Here, the linear tetrahedron method was employed with an 8 × 8 × 6 k-mesh. Results obtained from the WIEN2k code were found to agree well with those from the FPLO code.

## 3. RESULTS AND DISCUSSION

The crystal structure of $Ba_2NiOsO_6$ was refined by the Rietveld technique on an SXRD profile collected at room temperature. The SXRD pattern is shown in Figure 1, and the refined crystallographic parameters are listed in Table 1. The results indicate that $Ba_2NiOsO_6$ crystallizes in a fully ordered double perovskite structure, which is cubic with space group *Fm-3m* (similar to many other DPOs).[37, 38] The degree of order of the Ni and Os atoms was carefully investigated in the refinement procedure. First, the atoms were assumed to occupy the 4*a* and 4*b* crystallographic sites randomly and partially at equimolar ratio. Then, a fully ordered model, in which Ni remained solely at the 4*b* site and Os at the 4*a* site, was tested. The fully ordered model was found to provide the best refinement quality. We therefore conclude that $Ba_2NiOsO_6$ has fully ordered Ni and Os atoms, similar to $Sr_2NiOsO_6$ and $Ca_2NiOsO_6$.[39] The order has been attributed to distinctive differences between the charge, size, and electronic configuration of the Ni and Os atoms.[38, 39]

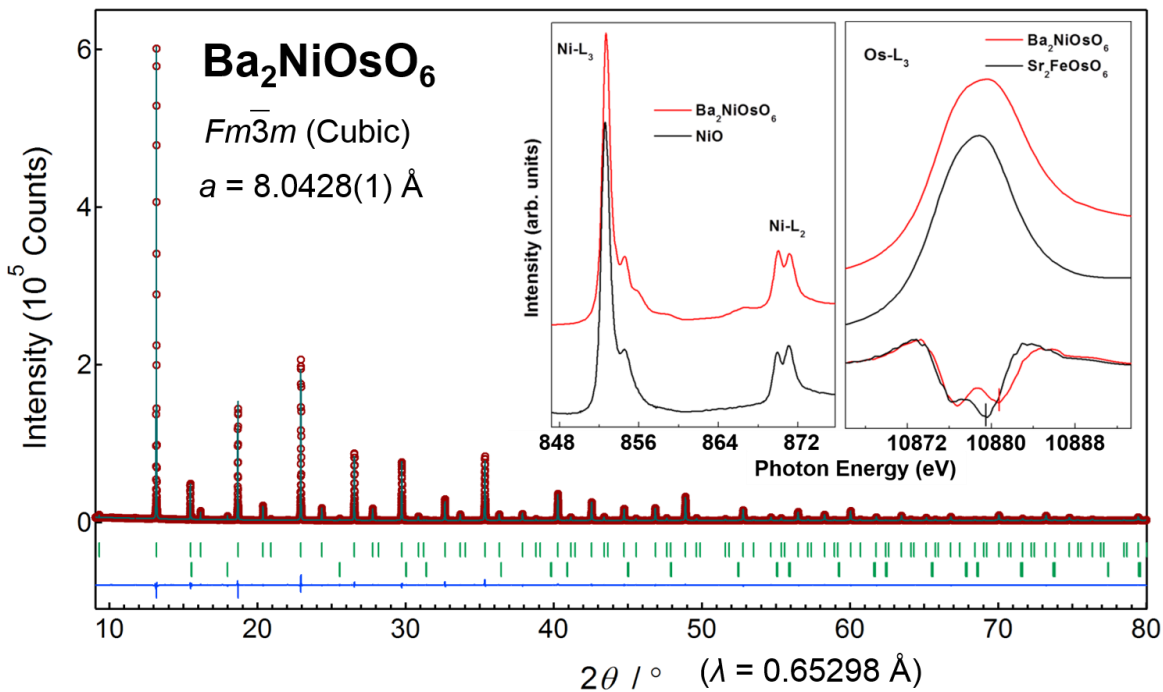

Figure 1. The Rietveld refinement of the SXRD profile collected at room temperature for $Ba_2NiOsO_6$. The expected Bragg reflections are marked by ticks for $Ba_2NiOsO_6$ (top) and NiO (bottom). The estimated mass proportion was $Ba_2NiOsO_6$:NiO = 0.976:0.024. The insets depict the room temperature Ni-$L_{2,3}$ XAS spectra of $Ba_2NiOsO_6$ and NiO ($Ni^{2+}$ reference) and the Os-$L_3$ XAS spectra of $Ba_2NiOsO_6$ and $Sr_2FeOsO_6$ ($Os^{5+}$ reference) and their second derivative.

**Table 1. Atomic positions and thermal parameters of $Ba_2NiOsO_6$**

| Atom | site | Occupancy | x | y | z | $B$ (Å$^2$) |
|---|---|---|---|---|---|---|
| Ba | 8c | 1 | 0.25 | 0.25 | 0.25 | 0.36(6) |
| Os | 4a | 1 | 0 | 0 | 0 | 0.11(4) |
| Ni | 4b | 1 | 0.5 | 0 | 0 | 0.19(2) |
| O | 24e | 1 | 0.2415(3) | 0 | 0 | 0.40(2) |

Note: Space group: $Fm$-$3m$; lattice parameter $a$ = 8.0428(1) Å, $Z$ = 4; $d_{cal}$ = 7.9102 g/cm$^3$. $R$ indexes were $R_p$ = 2.82% and $R_{wp}$ = 3.98%.

The local coordination at the magnetic Ni and Os ions was investigated. The Os–O bond length [1.943(2) Å] was similar to values of 1.948 and 1.955 Å for the $Os^{6+}$–O bond in $Ba_2CaOsO_6$ [40] and $Ba_2CuOsO_6$,[41] respectively, rather than 1.963 Å for the $Os^{5+}$–O bond in $Ba_2YOsO_6$.[42] This suggests that Os is nearly hexavalent. The bond valence sum (BVS) of Os was 5.72,[43] which also suggests hexavalency. The Ni–O bond length was 2.078(2) Å, which is comparable to 2.097 Å for the $Ni^{2+}$–O bond in $Ba_2NiWO_6$ [44] and distinctively different from 1.960 and 1.976 Å for the $Ni^{3+}$–O bond in $LaNiO_3$ [45] and $LiNiO_2$,[46] respectively. (We could not find any references for $Ba_2Ni^{3+}MO_6$ in accessible databases or the literature). The BVS of Ni was 2.07, which indicates that Ni is nearly divalent. To ensure charge compensation in bulk, the valence distribution should be nominally +2 and +6 for Ni and Os, respectively. Therefore, the formal electronic configurations should be $3d^8$ ($t_{2g}^6 e_g^2$; $S$ = 1) for Ni, provided that the exchange splitting is smaller than the crystal field splitting (CFS), and $5d^2$ ($t_{2g}^2 e_g^0$; $S$ = 1) for Os. (Here and throughout the paper, we use the term CFS in the usual meaning that includes both electrostatic and hybridization effects.) The parameters $R_0(Ni^{2+})$ = 1.675 Å, $R_0(Os^{6+})$ = 1.925 Å, and $B$ = 0.37 were used in the BVS calculations.[43]

To further examine the valence states of Ni and Os in $Ba_2NiOsO_6$, we measured the Ni-$L_{2,3}$ and Os-$L_3$ XAS spectra together with those of reference materials (see the insets in Figure 1). The identical energy positions and analogous multiplet spectral structures of the Ni-$L_{2,3}$ edges of $Ba_2NiOsO_6$ and NiO demonstrate the same $Ni^{2+}$ valence state and $NiO_6$ local symmetry in both compounds. In contrast, a spectral shift at the Os-$L_3$ edge of approximately 1.0(1) eV to higher energy from $Sr_2FeOsO_6$ to $Ba_2NiOsO_6$ indicates an increased valency of one from $Os^{5+}$ to $Os^{6+}$ in the latter.[47-50] The energy shift, the CFS, and the relative $t_{2g}$ and $e_g$ related spectral weights can be clearly discerned in the second derivative spectra at the bottom of the right inset in Figure 1. The $e_g$ related peak is shifted by 1.3(1) eV to higher energy in $Ba_2NiOsO_6$ relative to $Sr_2FeOsO_6$, and the spectral weight of the $t_{2g}$ related peak increases reflecting a smaller number of $t_{2g}$ holes[47] on $Os^{5+}$ ($t_{2g}^3$) compared to $Os^{6+}$ ($t_{2g}^2$). The second derivative spectra show also a larger CFS for $Os^{6+}$ in $Ba_2NiOsO_6$ compared with that of $Os^{5+}$ in $Sr_2FeOsO_6$ as found in observations of Ir oxides.[47]

We have compared the crystal structure of $Ba_2NiOsO_6$ with those of $A_2NiOsO_6$ ($A$ = Ca, Sr).[39] The crystal structure symmetry changes from monoclinic ($P2_1/n$; $A$ = Ca) to cubic ($Fm$-$3m$; Ba) via a tetragonal ($I4/m$; Sr) symmetry, as shown in Figure 2. An analogous symmetry change has been observed for $A_2FeMoO_6$ ($A$ = Ca, Sr, Ba),[37] which suggests that the size of $A$ plays a role in determining the lattice symmetry. The angle of the magnetic Os–O–Ni bond is altered from 151° ($Ca_2NiOsO_6$, monoclinic) [39] to 180° ($Ba_2NiOsO_6$, cubic) via 166°/180° ($Sr_2NiOsO_6$, tetragonal).[39] The angle variation seems to have a large impact on the magnetic properties. For example, $Ca_2NiOsO_6$ is canted AFM below 175 K, whereas $Sr_2NiOsO_6$ is AFM below 50 K. When the size of $A$ increases, the magnitude of the FM interaction increases as evidenced by the increase in Weiss temperature from 27 K in $Sr_2NiOsO_6$ (see Table S1) to 113(1) K in $Ba_2NiOsO_6$ (see the inset of Figure 3a). The Curie constant deduced from the 50 kOe FC curve for $Ba_2NiOsO_6$ is 1.48(2) emu mol$^{-1}$ K$^{-1}$, which corresponds to an effective moment ($\mu_{eff}$) of 3.46(2) $\mu_B$ per formula unit (f.u.). The Weiss temperature and the Curie constant deduced from the 0.1 kOe FC curve were 121(2) K and 1.28(2) emu mol$^{-1}$ K$^{-1}$ [$\mu_{eff}$ = 3.21(2) $\mu_B$ per f.u.], respectively, consistent with the absence of a marked change.

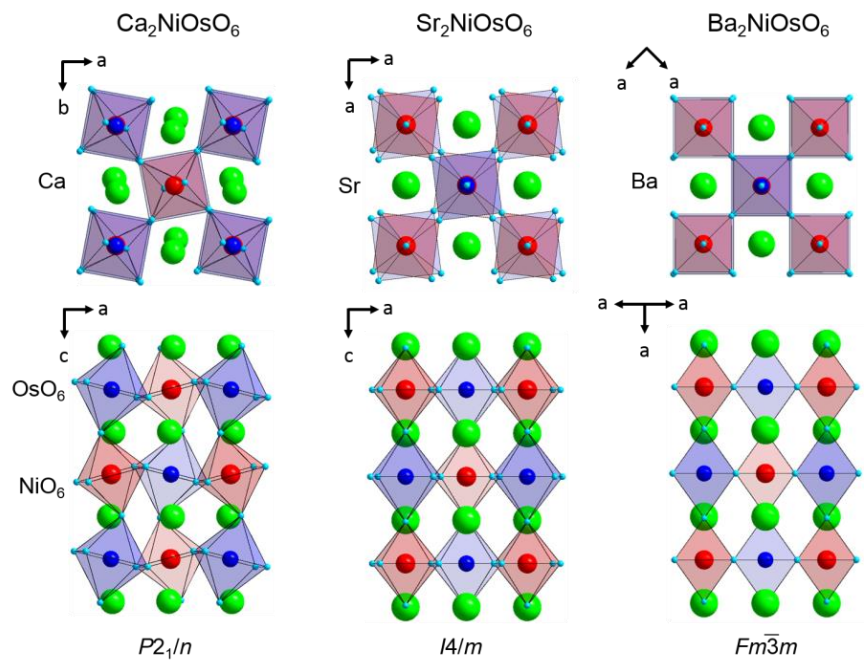

Figure 2. Structural comparison of the series $A_2NiOsO_6$ ($A$ = Ca, Sr, Ba). The crystal structure symmetry changes from monoclinic ($P2_1/n$) to cubic ($Fm$-$3m$) via a tetragonal ($I4/m$) symmetry. Green balls represent $A$ atoms; reddish and bluish octahedra represent $NiO_6$ and $OsO_6$.

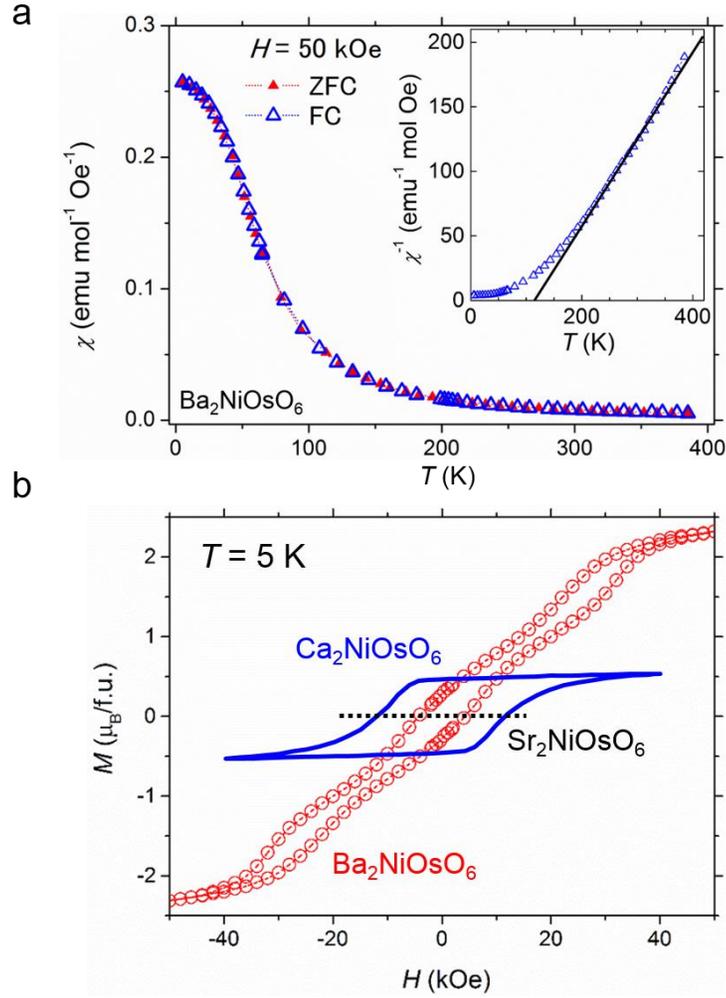

Figure 3. (a) Temperature dependence of the magnetic susceptibility of $Ba_2NiOsO_6$ measured at 50 kOe. The inset shows the inverse magnetic susceptibility. The solid line is the Curie-Weiss fit to the curve. (b) Isothermal magnetization loop of $Ba_2NiOsO_6$ at 5 K in comparison with the loops for $Ca_2NiOsO_6$ (taken from Ref. 39) and $Sr_2NiOsO_6$ (estimated from the magnetic susceptibility data in Ref. 39).

Although the FM interaction is large in $Ba_2NiOsO_6$, an AFM-like peak was observed in both the ZFC and FC curves at ~32 K ($T_{mag}$) (Figure S1). However, no noticeable peak is present in the 50 kOe curves (Figure 3a), which suggests a possible change of the magnetic order. To investigate the magnetism further, we conducted an isothermal magnetization measurement at a temperature below $T_{mag}$. The magnetization of $Ba_2NiOsO_6$ reaches approximately 2.3 $\mu_B$/f.u. at 50 kOe (at 5 K, Figure 3b). Intriguingly, the magnetic loop indicates a noticeable metamagnetic transition at approximately ±21 kOe for $Ba_2NiOsO_6$. For comparison, magnetic loops collected at 5 K for $Sr_2NiOsO_6$ and $Ca_2NiOsO_6$ are shown to emphasize the anomalous behavior of $Ba_2NiOsO_6$. A similar metamagnetic transition was observed for another DPO, $R_2CoMnO_6$ ($R$ = rare earth element). However, this was argued to be caused by the presence of significant antisite disorder.[51-54] Since such disorder is absent in $Ba_2NiOsO_6$, the origin of metamagnetism of $Ba_2NiOsO_6$ should be different.

We also conducted specific heat measurements near $T_{mag}$ with and without application of a magnetic field (Figure S2). The zero-magnetic field curve shows a broad cusp at ~32 K, whereas the cusp-like top shifts slightly toward lower temperature in an applied field of 10 kOe. In magnetic fields of 30 and 50 kOe, the cusp disappears completely. The disappearance is likely connected to the absence of an AFM-like peak on the $\chi$–$T$ curves at 50 kOe.

The temperature dependence of the electrical resistivity of polycrystalline $Ba_2NiOsO_6$ is shown in Figure 4. At room temperature the value is ~120 $\Omega$ cm, which is more than three orders of magnitude greater than the expected value for a metallic polycrystalline oxide. Electrical resistivity increased upon cooling and exceeded the instrumental limit at temperatures below 105 K indicating a semiconductor-like behavior. To estimate the activation energy of conduction, data points were plotted on a logarithmic scale as a function of inverse temperature and fit to the Arrhenius equation. The fit provided an activation energy of 0.31 eV (see the inset to Figure 4). In a field of 50 kOe, the electrical resistivity remained high and exceeded the instrumental limit at low temperatures as in zero field (see the plot of solid circles).

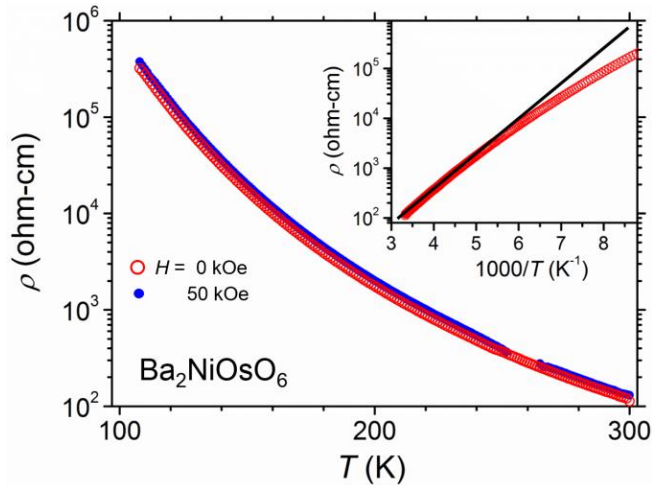

Figure 4. The temperature dependence of $\rho$ for polycrystalline $Ba_2NiOsO_6$. The inset shows the Arrhenius plot of the data and the fitting line, which yield an activation energy of 0.31 eV. Data measured in a magnetic field of 50 kOe also are presented.

The magnetic order of $Ba_2NiOsO_6$ was investigated by ND at low temperatures. Figures 5a and 5b show the ND patterns ($\lambda$ = 1.54 Å) collected at 50 and 4 K, respectively. The nuclear lattice was well refined by the room temperature lattice model both above and below the AFM-like peak temperature (~32 K), which indicates the absence of a lattice symmetry change down to 5 K regardless of the presence of the magnetic transition. Although the electronic configuration of $Os^{6+}$ (5d $t_{2g}^2$) implies a possible orbital order below $T_{mag}$,[55, 56] any indication of reduced lattice symmetries lower than cubic was undetected within the instrument resolution. For example, a refinement attempt with a tetragonal space group of $I4/m$ didn't converge to stable lattice parameters. A robust cubic symmetry lattice, which argues against orbital ordering, was found in a similar $Os^{6+}$ compound, $Ba_2CaOsO_6$.[57] The low temperature crystallographic parameters are summarized in Table S2.

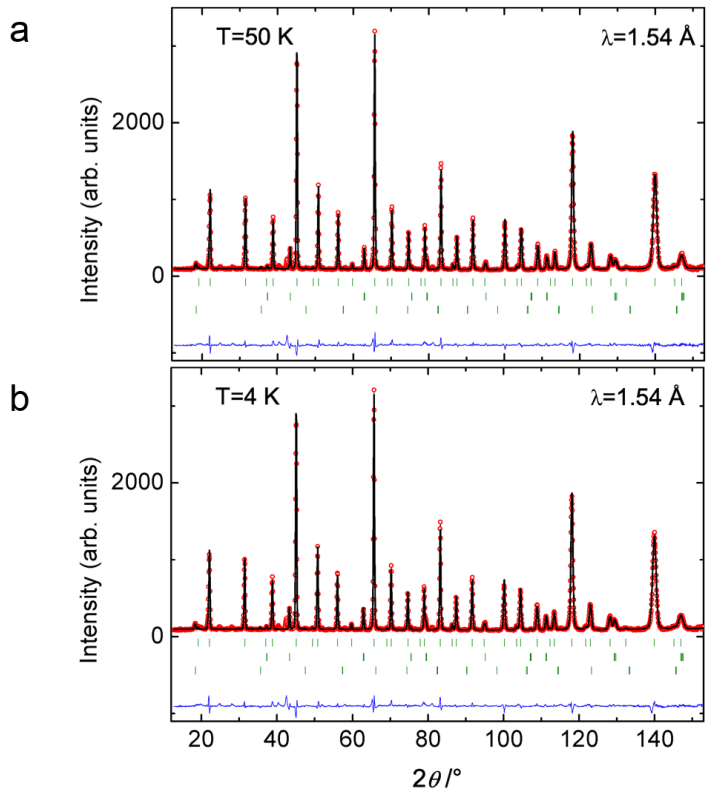

Figure 5. Rietveld refinement of the ND profiles ($\lambda$ = 1.54 Å) of $Ba_2NiOsO_6$ collected at temperatures of (a) 50 K and (b) 4 K. The expected Bragg reflections are marked by ticks for $Ba_2NiOsO_6$ (top), NiO (middle), and the magnetic lattice of NiO (bottom).

After changing $\lambda$ to 2.41 Å, we conducted ND measurements to investigate the magnetic order. The difference between the profiles at 5 and 75 K clearly revealed the presence of modulated AFM order as shown in Figure S4 (detailed analysis is provided in the supplemental material). The modulated AFM order transformed to a collinear FM order in a magnetic field above 21 kOe at 5 K. For example, the difference between the zero-field and non-zero-field ND profiles clearly revealed the presence of FM order (Figure 6a), which gradually developed with increasing magnetic field (Figure 6b). Figure 6c shows the applied magnetic field dependence of the magnetic peak intensity at $2\theta = 30°$, which indicates that the critical field ($H_c$) is 21 kOe at 5 K. The magnetic field observations throughout the ND measurements accord well with the metamagnetic transition found in the isothermal magnetization measurements. The FM order also was refined by analysis of the ND profile collected at 45 kOe. All Ni and Os magnetic moments lie along the [100] or equivalent [010] or [001] direction (Figure 6d) indicating the common plane of the modulated AFM and FM orders.

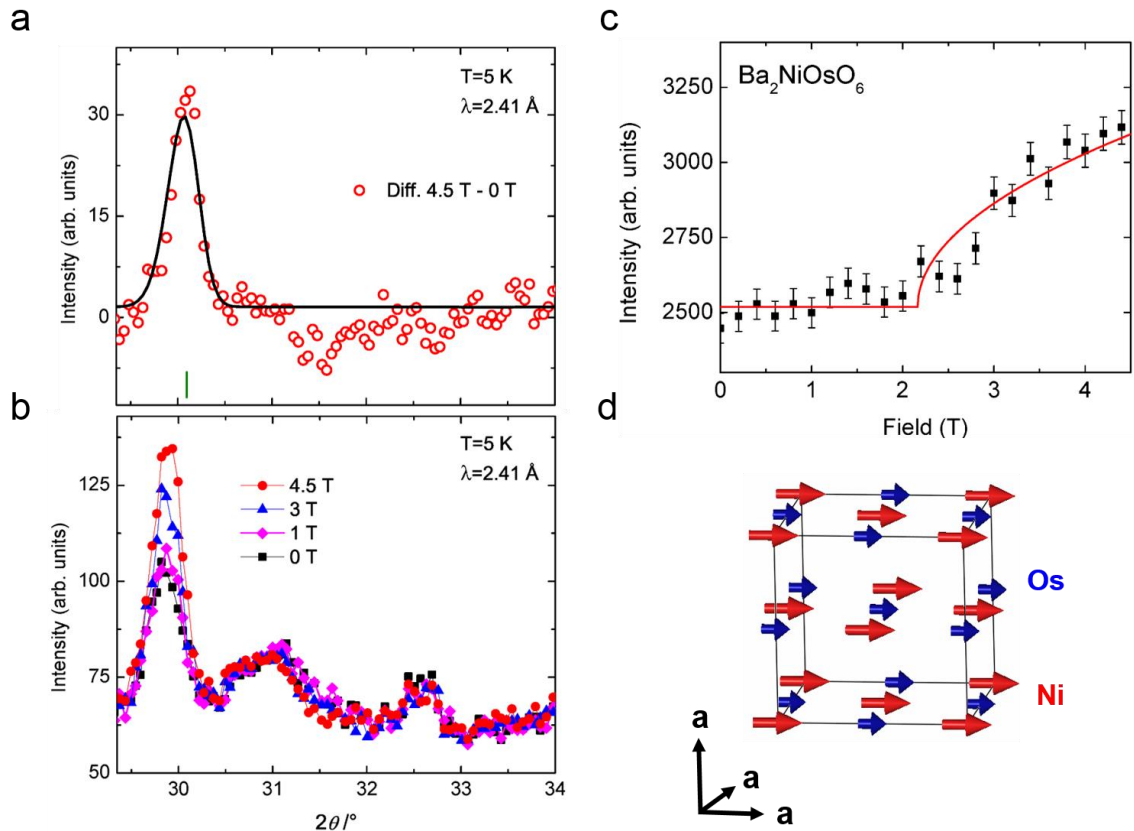

Figure 6. (a) Difference between the zero-field profile and the profile collected in a magnetic field of 45 kOe at 5 K. (b) Evolution of a magnetic peak with increasing magnetic field at 5 K. (c) Magnetic field dependence of the intensity of the magnetic peak at $2\theta = 30°$ ($\lambda = 2.41$ Å) at a fixed temperature of 5 K. The solid line and curve are a guide for the eyes. (d) Magnetic order model of the FM state of $Ba_2NiOsO_6$ depicted from the analysis of the ND profiles. The magnetic easy axis lies in the cubic [100], [010], or [001] direction.

The magnetic moments in the FM state at 5 K were estimated to be 2.13(8) and 0.97(5) $\mu_B$ per Ni and Os, respectively. Although the Ni moment is comparable to the spin only moment estimated from a simplified model, the Os moment is significantly smaller. This is most likely due to the impact of the SOC of Os, as suggested in Ref. 58, and to the extended orbitals of the 5d atom leading to increased hybridization, as suggested in Ref. 59. The magnetization of $Ba_2NiOsO_6$ at 5 K and 50 kOe is approximately 2.3 $\mu_B$/f.u. (Figure 3b), although the ND analysis yields approximately 3 $\mu_B$/f.u. in total. This indicates incomplete saturation in a 50 kOe field, which is evident from the relatively large gradient, *dM/dH*, at this strength (see Figure 3b). Further high field magnetization studies may confirm the ND result.

We have analyzed the electronic and magnetic states of $Ba_2NiOsO_6$ by DFT. We calculated the total energies of supercells with magnetic alignments corresponding to one FM, two AFM, and three FIM orderings (see Figure 7). Using up/down spins to indicate the magnetic moment on the Ni1, Ni2, Os1, and Os2 atoms, the orderings investigated are FM1-↑↑↑↑, AF1-↑↓↑↓, AF2-↑↓↓↑, FI1-↑↑↓↓, FI2-↑↓↑↑, and FI3-↑↑↑↓, respectively. These arrangements are considered in tetragonal supercells (space group 123, *P*4/mmm) containing two chemical unit cells. We double checked the calculations using orthorhombic supercells (space group 47, *P*mmm) with four chemical unit cells and pseudo-cubic lattice parameters $a = b = c$. The two settings produced nearly identical results (with differences below 4 meV/f.u.) for calculations conducted without SOC. With SOC, the total energies obtained with the tetragonal setting did not fulfill the expected symmetry relations. Thus, we report only results obtained with the orthorhombic setting. The latter was double checked by performing GGA + *U* + SOC calculations in a cubic cell with one chemical unit for the FM1 case. The magnetic anisotropy energies were 18 meV/f.u. (orthorhombic setting, see Table 2) and 13 meV/f.u. (cubic setting). From this comparison and the small differences between the orthorhombic and tetragonal settings, we assume that the other energy differences in Table 2 have a similar precision of ca. 5 meV/f.u.

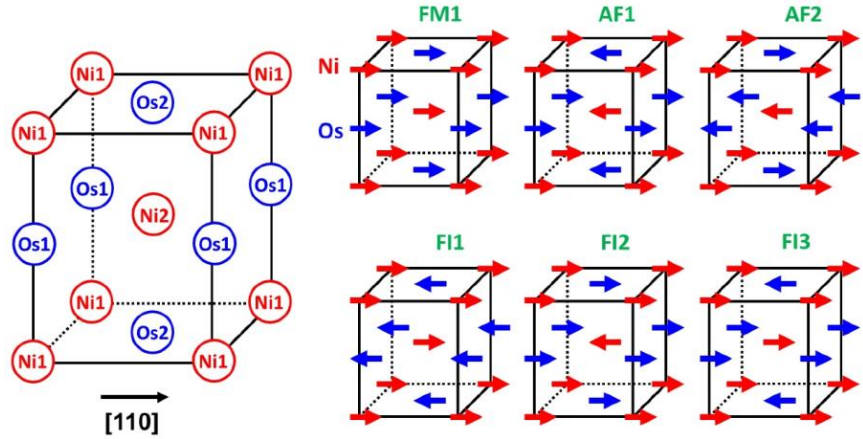

Figure 7. Left: Sketch of the tetragonal supercell (space group 123, $P4/mmm$) used for DFT calculations of possible magnetic arrangements. The cubic [110] direction is indicated. Right: Schematic of the magnetic arrangements. Red/blue arrows indicate magnetic moments on the Ni/Os atoms. The magnetic easy axis was found to lie along the cubic [110] direction for FM1 and AF2 by GGA+$U$+SOC.

Table 2. Relative total energies (meV/f.u.) calculated within GGA, GGA + $U$, and GGA + $U$ + SOC schemes for $Ba_2NiOsO_6$. For the latter scheme, the orientations of the magnetic moments are given with respect to the cubic axes. Results for the [110] orientation were obtained only for the three spin arrangements with the lowest energies in GGA or GGA + $U$.

|  | FM1 | AF1 | AF2 | FI1 | FI2 | FI3 |
| --- | --- | --- | --- | --- | --- | --- |
| GGA | 0 | 85 | 40 | 115 | 61 | 58 |
| GGA+$U$ | 0 | 73 | 45 | 82 | 58 | 40 |
| GGA+$U$+SOC [001] | 18 | 68 | 41 | 91 | 55 | 53 |
| GGA+$U$+SOC [100] | 18 | 48 | 25 | 91 | 37 | 52 |
| GGA+$U$+SOC [110] | 0 |  | 16 |  |  | 28 |

By comparing the total energies of the different magnetic arrangements, we find that the FM1 order has the lowest total energy regardless of the calculation scheme (Table 2). The energy difference to the next lowest order (AF2) is very small, ~16 meV per f.u. in GGA+$U$+SOC. The difference is almost comparable to the precision of the method, which implies a competition between the FM1 and AF2 orders. Other possible magnetic orders (not shown here) were found at much higher energies and were excluded from further investigation.

The FM1 alignment is identical to the experimental FM state in an external magnetic field (Figure 6d). Figure 8 shows the DOS of FM1 for cases (a) without and (b) with SOC. The band structures in Figures 9a and 9b show a metallic state for GGA+$U$ and a gapped state for GGA+$U$+SOC.

For a more detailed view, we considered the partial DOS (Figure S7). The Ni 3d band is fully occupied in one spin channel and partially occupied (nominally $t_{2g}^3 e_g^0$) in the other, both with and without SOC. Due to the combined effect of CFS and $U = 5$ eV, the unoccupied Ni $e_g$ states lie about 2 eV above $E_F$. The O 2p orbitals hybridize strongly with the Ni 3d orbitals below -1.5 eV (the valence region) and with the Os 5d orbitals.

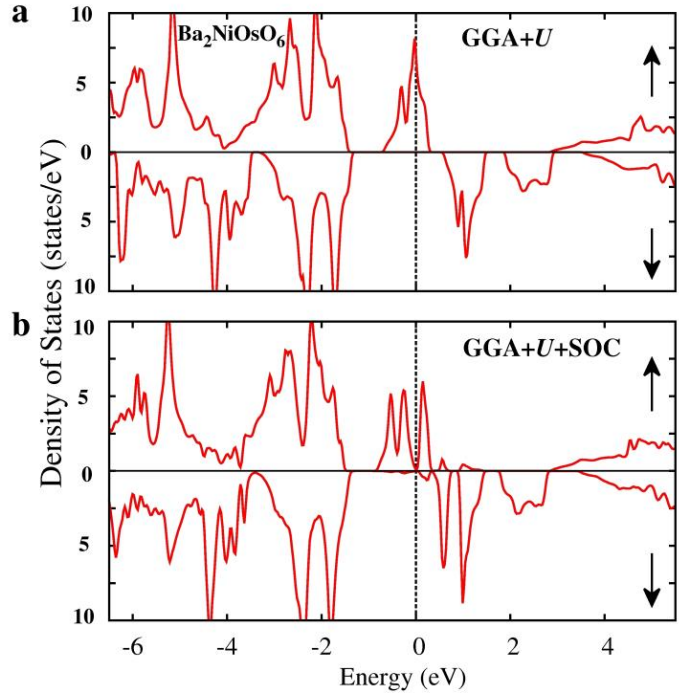

Figure 8. The total DOS (per f.u.) for spin-up (↑) and spin-down (↓) channels of the FM1 state of Ba$_2$NiOsO$_6$ obtained in the (a) GGA+$U$ and (b) GGA+$U$+SOC schemes. The vertical dotted line indicates $E_F = 0$.

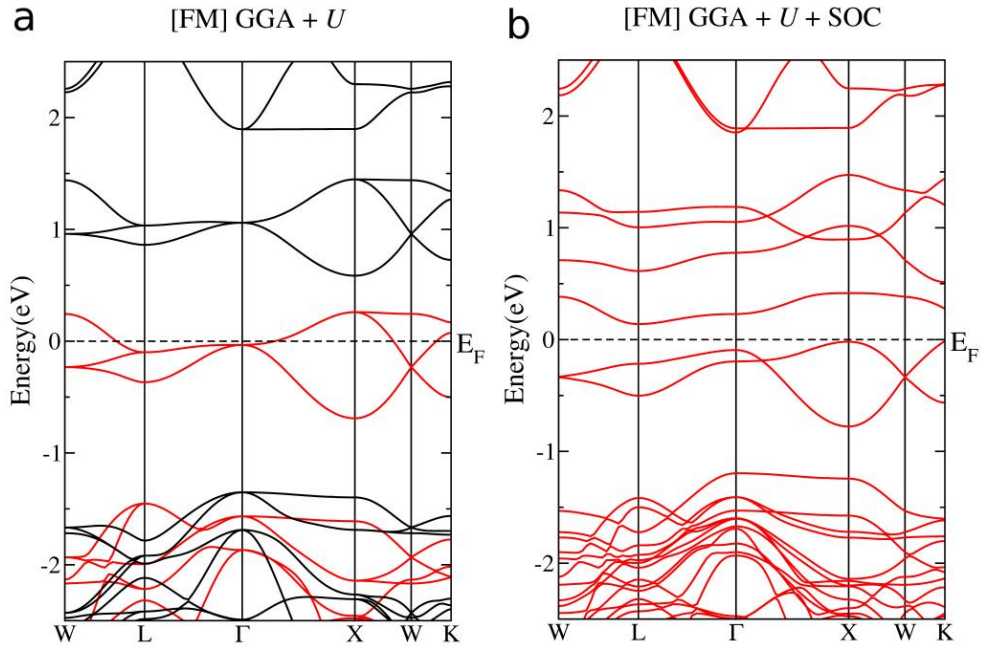

Figure 9. Band structures of the FM1 state of $Ba_2NiOsO_6$ within (a) GGA+$U$ (red: spin-up; black: spin-down) and (b) GGA+$U$+SOC. Dashed horizontal lines at zero correspond to $E_F$.

The most interesting band located at the Fermi level is composed mainly of Os 5d states (Figure S7). We performed a decomposition into $m_l$-states (Figure S8), which confirms the expected $t_{2g}$ character of the band. Without SOC (Figure 9a), the Os $t_{2g}$ states form two band complexes, each containing three individual bands: a spin-up complex around the Fermi level and a spin-down complex around 1 eV. The former is partially occupied with two electrons per Os. It cannot be split without breaking the non-relativistic cubic symmetry, such that application of $U$ does not open a gap. With SOC (Figure 9b), the $t_{2g}$ states are split into two $\Gamma_8$ states and one $\Gamma_7$ state. However, SOC alone is not sufficient to open a gap, because the band dispersion is greater than the spin-orbit splitting, which leads to an (indirect) overlap of the band states. Although we could not directly measure the strength of the SOC from the calculation, a comparison between the degree of splitting of the $t_{2g}$ band within the GGA and GGA + SOC schemes roughly estimates the spin-orbit splitting to be ~0.17 eV. Due to symmetry reduction upon inclusion of SOC, application of Coulomb corrections may increase the splitting between occupied and unoccupied bands, which will open the gap at a certain value of $U$. A schematic illustration of the mechanism is depicted in Figure 10. For the sake of clarity, details of the DOS are ignored or changed, and the oxygen DOS is omitted.

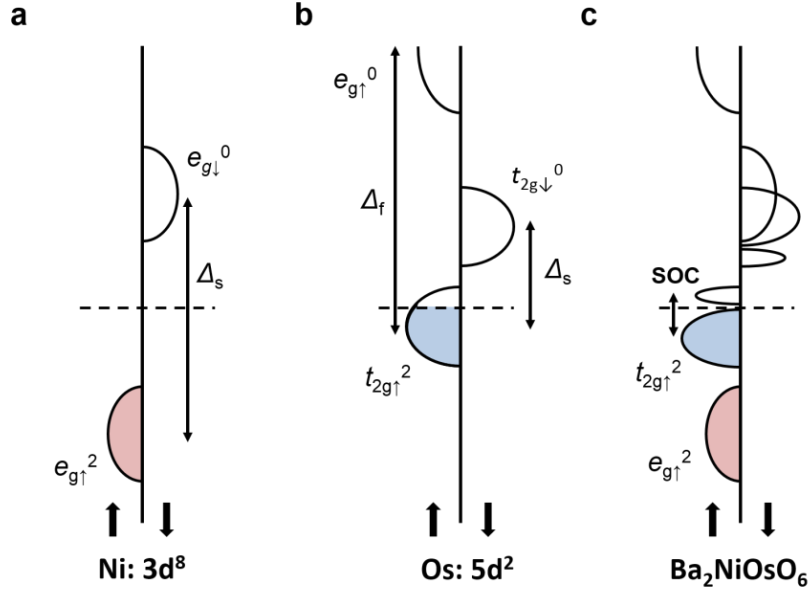

Figure 10. A schematic view of the d-band DOS structure of the FM1 state of $Ba_2NiOsO_6$. The partial DOS for (a) Ni 3d and (b) Os 5d bands and (c) the total DOS. The symbols $\Delta_s$ and $\Delta_f$ represent the spin exchange splitting and crystal field splitting (including correlation effects), respectively.

## 4. CONCLUSION

Cubic $Ba_2NiOsO_6$ was synthesized for the first time by solid-state reaction at 6 GPa and 1500 °C. SXRD and ND revealed cubic lattice symmetry (*Fm-3m*) with lattice parameter $a = 8.0428(1)$ Å at room temperature [8.0298(1) Å at 4 K]. XAS of the Ni and Os core levels indicated that the formal charges on Ni and Os are essentially +2 and +6, respectively. Thus, the formal electronic configurations should be $3d^8$ ($t_{2g}^6 e_g^2$; $S = 1$) for Ni and $5d^2$ ($t_{2g}^2 e_g^0$; $S = 1$) for Os. The combined evaluation of theoretical and experimental results establish $Ba_2NiOsO_6$ as an FM insulator with $T_{mag}$ of ~100 K. SOC plays an essential role in opening the charge gap. A metamagnetic transition was found at 5 K, where a modulated AFM state transforms to the FM state at a moderate magnetic field ($H_c$ ~21 kOe at 5 K) without altering the semiconducting property. DFT calculations suggest that FM and helical AFM orders have comparable total energies and are responsible for the metamagnetic transition. We conclude that both Coulomb and relativistic effects must be considered equally in $Ba_2NiOsO_6$. The situation is similar to the cases of the platinum group oxide $Sr_2IrO_4$[60-62] and α-$RuCl_3$,[63] although these are essentially AFM materials.

Recently, a similar material, $Ba_2NsOsO_6$, having $5d^1$ ($t_{2g}^1 e_g^0$) configuration of Os, has theoretically been studied, suggested that it is an uncommon FM insulator.[64] Although the Weiss temperature of $Ba_2NsOsO_6$ is weekly negative (-10 K), it might be the first material being clarified as a FM Dirac-Mott insulator ($T_{mag}$ = 6.8 K). The preset compound is likely the second case (or possibly later); however, the FM properties are dramatically improved as evidenced by such as the substantially positive Weiss temperature (+113 K) and increased $T_{mag}$ near 100 K. Thus, the newly developed cubic double perovskite could be useful to deepen understanding nature of FM Dirac-Mott insulator.

Cubic $Ba_2NiOsO_6$ should increase general interest in platinum group oxides. Platinum group metals and their complexes have been used in many practical applications, because of their excellent catalytic activities.[65] However, electrical properties and other practical features are not well known apart from high electrical conductivity and robust corrosion resistance. Recent studies of Ir and Os oxides have shown that the significant SOC and extended 5d orbitals in the oxide produce outstanding electric and magnetic properties as found in Slater insulators,[19, 66] Dirac-Mott insulators,[60, 62] and ferroelectric metals.[67] The largest reported spin-phonon-electron coupling was recently discovered in an osmium oxide, which further illustrates this potential of high-valent platinum group oxides.[68] Growing interest in novel physical properties, as exemplified by the FM semiconductivity of cubic $Ba_2NiOsO_6$, should motivate further study of platinum group oxides. Our report of cubic $Ba_2NiOsO_6$ heralds a new class of FM insulator oxides, which may be useful in developing a practical magnetic semiconductor that can be employed in spintronic and quantum magnetic devices.

**Acknowledgements**

The authors would like to thank the staff of BL15XU, NIMS, and SPring-8 for their help at the beamline. The SXRD measurements were performed under the approval of the NIMS Beamline Station (Proposal No. 2014A4504, 2014B4501, 2015A4502, and 2016B4504). The XAS measurements were supported by Chin-Wen Pao. M.P.G thanks the Alexander von Humboldt Foundation for financial support through the Georg Forster Research Fellowship Program. M.P.G. thanks K. Koepernik and R. Laskowski for helpful discussion, and M.R. thanks M. Knupfer, U. Rößler, and H. Rosner for helpful discussion. This research was supported in part by the World Premier International Research Center of the Ministry of Education, Culture, Sports, Science and Technology (MEXT) of Japan and the Japan Society for the Promotion of Science (JSPS) through a Grant-in-Aid for Scientific Research (25289233, 15K14133, 16H04501). The research conducted at ORNL's High Flux Isotope Reactor was sponsored by the Scientific User Facilities Division, Office of Basic Energy Sciences, U.S. Department of Energy.

# Supplemental Material

# Ba$_2$NiOsO$_6$: a Dirac-Mott insulator with ferromagnetism near 100 K

Hai L. Feng,[a,b,*] Stuart Calder,[c] Madhav Prasad Ghimire,[d,e,*] Ya-Hua Yuan,[a,f] Yuichi Shirako,[g,#] Yoshihiro Tsujimoto,[a] Yoshitaka Matsushita,[h] Zhiwei Hu,[b] Chang-Yang Kuo,[b] Liu Hao Tjeng,[b] Tun-Wen Pi,[i] Yun-Liang Soo,[i,j] Jianfeng He,[a,f] Masahiko Tanaka,[k] Yoshio Katsuya,[k] Manuel Richter,[d,l] Kazunari Yamaura [a,f]

**Table S1.** List of double perovskite oxides, $A_2M$OsO$_6$, in which $A$ = Ca, Sr, Ba, and $M$ = a 3d-transition metal.

| Composition | Valence of $M$/Os and $d$ electron count | S.G.[a] above $T_{mag}$[b] | $T_{mag}$/K | Curie-Weiss parameters | | Note | Refs. |
|---|---|---|---|---|---|---|---|
| | | | | $\Theta_w$/K | $\mu_{eff}/\mu_B$ | | |
| Sr$_2$CrOsO$_6$ | 3+/5+  3d$^3$–5d$^3$ | $Fm$-3$m$ | 725 | | | Asymmetric AFM | S1-3 |
| Ca$_2$FeOsO$_6$ | 3+/5+  3d$^5$–5d$^3$ | $P2_1/n$ | 320 | | | FIM | S4-7 |
| Sr$_2$FeOsO$_6$ | 3+/5+  3d$^5$–5d$^3$ | $I4/m$ | 140, 67 | +80 | 4.32 | Competing AFM | S4, 8, 9 |
| Ca$_2$CoOsO$_6$ | 2+/6+  3d$^7$–5d$^2$ | $P2_1/n$ | 145 | +5 | 4.16 | FIM | S10 |
| Sr$_2$CoOsO$_6$ | 2+/6+  3d$^7$–5d$^2$ | $I4/m$ | 108, 70 | −51 | 4.46 | Two AFM transitions | S11, 12 |
| Ca$_2$NiOsO$_6$ | 2+/6+  3d$^8$–5d$^2$ | $P2_1/n$ | 175 | | | Canted AFM | S13 |
| Sr$_2$NiOsO$_6$ | 2+/6+  3d$^8$–5d$^2$ | $I4/m$ | 50 | +27 | 3.44 | AFM | S13-15 |
| Ba$_2$NiOsO$_6$ | 2+/6+  3d$^8$–5d$^2$ | $Fm$-3$m$ | 100, 32 | +113 | 3.46(2) | FM semiconductor | This work |
| Sr$_2$CuOsO$_6$ | 2+/6+  3d$^9$–5d$^2$ | $I4/m$ | 20 | −40 | 2.07 | AFM | S14, 16, 17 |
| Ba$_2$CuOsO$_6$ | 2+/6+  3d$^9$–5d$^2$ | $I4/m$ | 70 | −13(3) | 2.33(2) | 2-dimensional AFM | S18 |

[a] Space group; [b] Magnetic transition temperature

**Supplemental description of double perovskite oxides:** Solid-state double perovskite oxides (DPOs) are potentially applicable for use in the anode of solid oxide fuel cells, with candidate materials including Sr$_2$Mg$_{1-x}$Mn$_x$MoO$_6$ [S19-21] and PrBaMn$_2$O$_{5+\delta}$.[S22] Other DPOs exhibiting a large spin polarization at the Fermi level and a high magnetic ordering temperature are expected to be useful as a practical spintronic material, such as Sr$_2$FeMoO$_6$.[S23-29] Much experimental progress has been made, particularly on the magnetic properties of 3d/5d(4d) DPOs and related materials.[S1-3, 6-12, 30-33] One major achievement in this field is the discovery of a high-temperature ferromagnetic (FM) transition at 635 K in the nearly half-metallic compound $Sr_2CrReO_6$.[S26] However, the spin polarization of this material turned out to be unsatisfactory for use in a practical device.[S25] Another major achievement is the prediction of the (likely novel) asymmetric antiferromagnetic (AFM) state [S2] for $Sr_2CrOsO_6$ with a magnetic transition temperature ($T_{mag}$) of 725 K.[S3] Currently, $Sr_2CrOsO_6$ has the highest known $T_{mag}$ among all the DPOs, initiating a number of theoretical investigations.[S1, 2, 30-33]

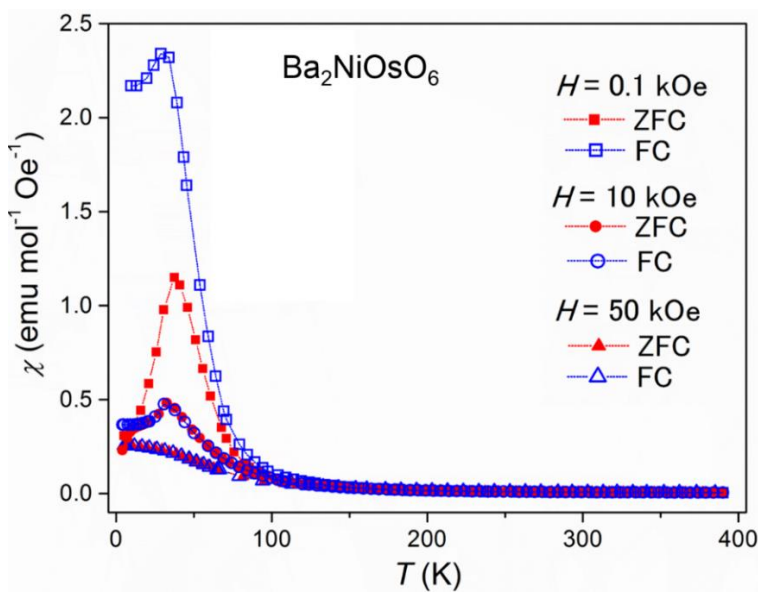

Figure S1. Temperature dependence of the magnetic susceptibility of $Ba_2NiOsO_6$ measured in various magnetic fields.

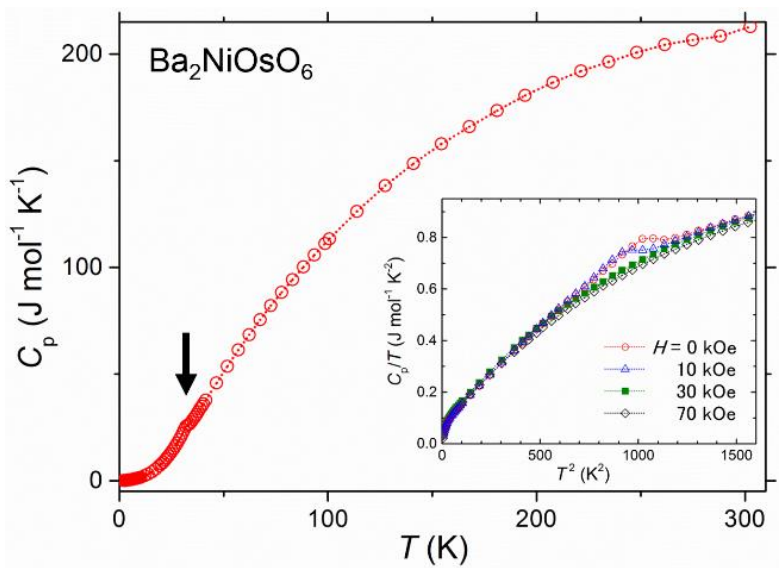

Figure S2. Specific heat of $Ba_2NiOsO_6$ measured with and without an applied magnetic field. The arrow indicates a cusp at ~32 K.

**Supplemental description of specific heat measurement:** We have roughly estimated the transition entropy, $\Delta S$, over the anomaly at $T_{mag}$ of ~32 K by subtracting the phonon contribution from the total $C_p$. The value of $\Delta S$ is approximately $0.1R$, where $R$ is the universal gas constant. This value corresponds to only 5% of the spin-only magnetic entropy of $2R\ln3$ for $Ni^{2+}$ ($e_g^2$) and $Os^{6+}$ ($t_{2g}^2$). The small magnetic entropy is likely caused by the presence of short-range magnetic order above the ordering temperature. Indeed, a FM-like short-range order seems to develop on cooling below ~100 K, as evidenced by the $\chi$–$T$ curves and the magnetization loops at 50 and 200 K (Figure S3).

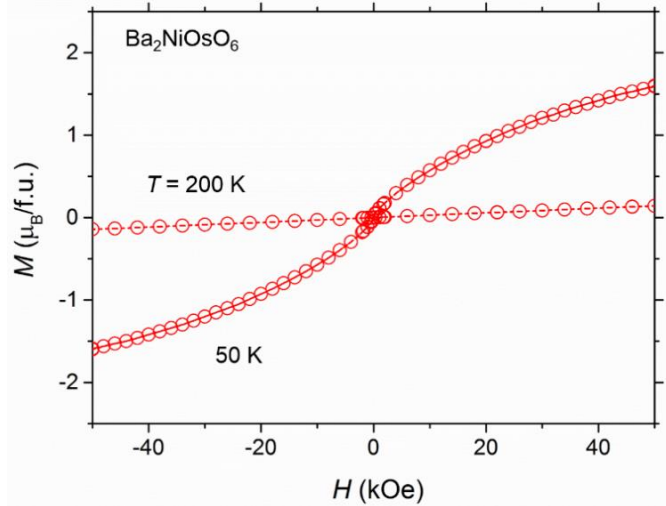

Figure S3. Magnetization loops of $Ba_2NiOsO_6$ measured at 50 and 200 K.

**Table S2.** Low-temperature crystallographic parameters of $Ba_2NiOsO_6$ determined by analysis of the ND profiles collected at 50 K and 4 K

| Atom | site | Occupancy | x | y | z | B (Å²) |
|---|---|---|---|---|---|---|
| $T$ = 50 K | | | | | | |
| Ba | 8c | 1 | 0.25 | 0.25 | 0.25 | 0.12(2) |
| Ni | 4a | 1 | 0 | 0 | 0 | 0.24(4) |
| Os | 4b | 1 | 0.5 | 0.5 | 0.5 | 0.15(4) |
| O | 24e | 1 | 0.2596(2) | 0 | 0 | 0.29(2) |
| $T$ = 4 K | | | | | | |
| Ba | 8c | 1 | 0.25 | 0.25 | 0.25 | 0.10(3) |
| Ni | 4a | 1 | 0 | 0 | 0 | 0.25(4) |
| Os | 4b | 1 | 0.5 | 0.5 | 0.5 | 0.12(4) |
| O | 24e | 1 | 0.2600(2) | 0 | 0 | 0.27(2) |

Note: Space group: $Fm$-$3m$; the lattice parameter $a$ = 8.0303(1) Å at 50 K and 8.0298(1) Å at 4 K, $Z$ = 4; $d_{cal}$ = 7.9467 g/cm³ (50 K) and 7.9483 g/cm³ (4 K). $R$ indexes were $R_p$ = 2.20% and $R_{wp}$ = 3.37% (50 K); and $R_p$ = 2.40% and $R_{wp}$ = 3.69% (4 K).

**Supplemental description of neutron diffraction study:** The difference between the profiles (at 5 and 75 K) clearly revealed the presence of magnetic peaks (see the peaks at $2\theta$ of ~31.5° and 32.8° in Figure S4a). Figure S4b shows the temperature dependence of the intensity of the magnetic peak at $2\theta$ of ~31.5°, indicating the establishment of long-range order at a temperature consistent with the anomaly in susceptibility and the zero-field anomaly in specific heat. The magnetic reflections measured by ND reveal AFM long range ordering and rule out FM ordering at $T_{mag}$. The only magnetic propagation vector to satisfy all reflections was (0, 0, 1/8). Following a representational analysis approach using this propagation vector with Os at the (0, 0, 0) site and Ni at the (0.5, 0.5, 0.5) site yields a decomposition of the magnetic representation ($\Gamma_{mag}$) into the following irreducible representations (IR): $\Gamma_{mag} = 0\Gamma_1^1 + 1\Gamma_2^1 + 0\Gamma_3^1 + 0\Gamma_4^1 + 1\Gamma_5^2$, following the numbering scheme of Kovalev.[S34] The basis vectors for the non-zero IRs ($\Gamma_2$ and $\Gamma_5$) are shown in Table S3. $\Gamma_2$ does not reproduce scattering at a $2\theta$ of 32.8° since this corresponds to a (0, 0, $L$) reflection and so can be discounted. Instead $\Gamma_5$ accurately models the reflections observed with ND. Considering only the real basis vector components gives an AFM ordered structure shown in Figure S5, in which the FM planes are antiferromagnetically coupled with the neighbor planes with a period of 8 nuclear unit cells along one of the cubic directions, forming a sine wave-like AFM spin density wave order over a long range. This magnetic structure is, however, at odds with the insulating behavior observed from resistivity that points to local moments and not varying moments. Therefore we took into account the imaginary basis vector components that also form part of this IR. The solution gives equivalent magnetic scattering as the spin density wave structure, however in this case the magnetic structure is described by local moments of fixed amplitude that form a helical structure in which the magnetic moments of Ni and Os in a {001} plane are ordered ferromagnetically in the plane (Figure S6a). The FM planes are coupled with the neighbor planes with a rotation periodicity of 8 nuclear unit cells.

The magnitude of the magnetic moments of Ni and Os in the modulated AFM state at 5 K were estimated from the analysis of the ND patterns by scaling the magnetic reflection intensities for the modulated AFM order to the nuclear structural reflections and found to be 1.2(2) $\mu_B$ and 0.3(2) $\mu_B$ per Ni and Os, respectively. This is smaller than the spin-only moment of 2 $\mu_B$ for each ($e_g^2$ or $t_{2g}^2$). Because they are possibly influenced by other properties, such as hybridization, magnetic frustration, magnetic disorder,[S35] thermal fluctuation, and SOC, further studies are needed to understand the underlying mechanisms of the reduced moments.

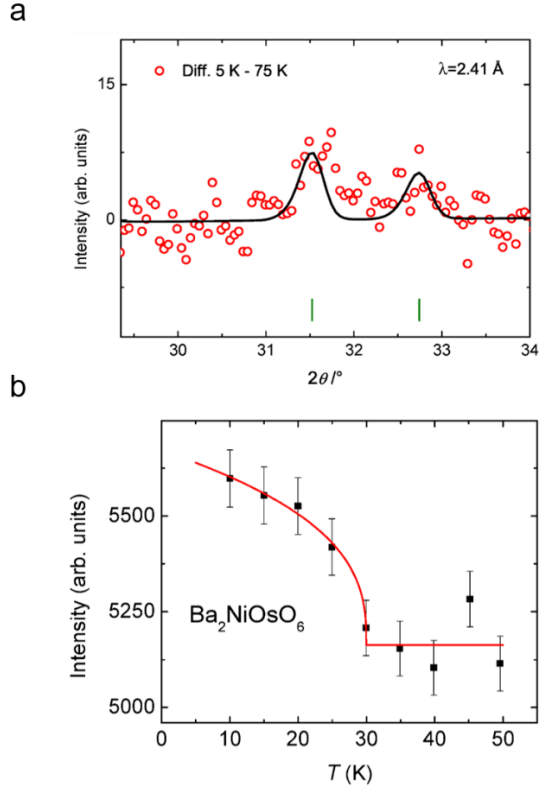

Figure S4. (a) Difference between the ND profiles collected at 5 K and 75 K without applying a magnetic field. (b) Temperature dependence of the intensity of a magnetic peak at $2\theta$ of ~31.5° of the ND profile ($\lambda = 2.41$ Å) collected in zero magnetic field.

**Table S3. Irreducible representations (IR) and Basis vectors (BV) for the space group *Fm*-3*m* with propagation vector *k* = (0, 0, 1/8). Both Os and Ni atoms have the same decomposition into $\Gamma_{mag}$ = $0\Gamma_1^1 + \Gamma_2^1 + 0\Gamma_3^1 + 0\Gamma_4^1 + \Gamma_5^2$.**

| IR | BV | Real BV components | | | Imaginary BV components | | |
|---|---|---|---|---|---|---|---|
| | | $m_a$ | $m_b$ | $m_c$ | $im_a$ | $im_b$ | $im_c$ |
| $\Gamma_2$ | $\Psi_1$ | 0 | 0 | 1 | 0 | 0 | 1 |
| $\Gamma_5$ | $\Psi_2$ | 1 | 0 | 0 | 1 | 0 | 0 |
| | $\Psi_3$ | 0 | -1 | 0 | 0 | -1 | 0 |

S5

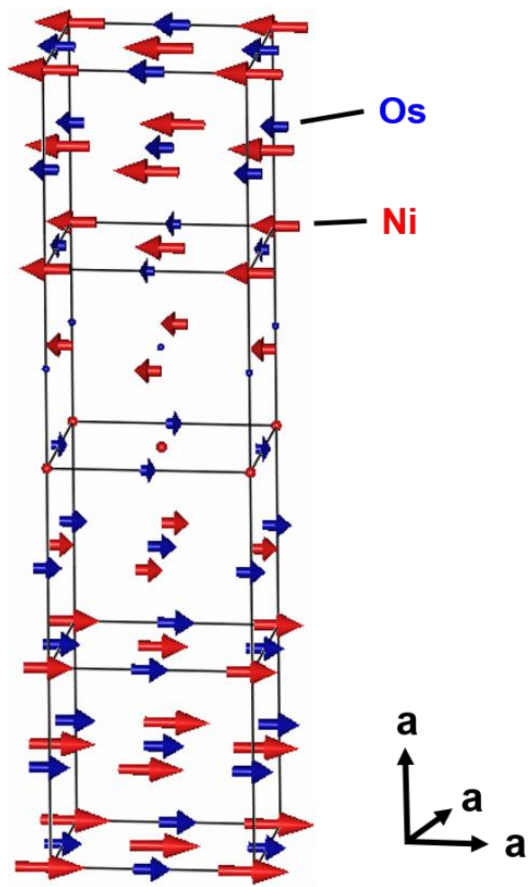

Figure S5. A sine wave-like model of the AFM state of $Ba_2NiOsO_6$, depicted from the analysis of the ND profiles. The AFM unit cell consists of 8 nuclear unit cells along one of the cubic lattice axes. The magnetic moments are aligned perpendicular to the propagation vector.

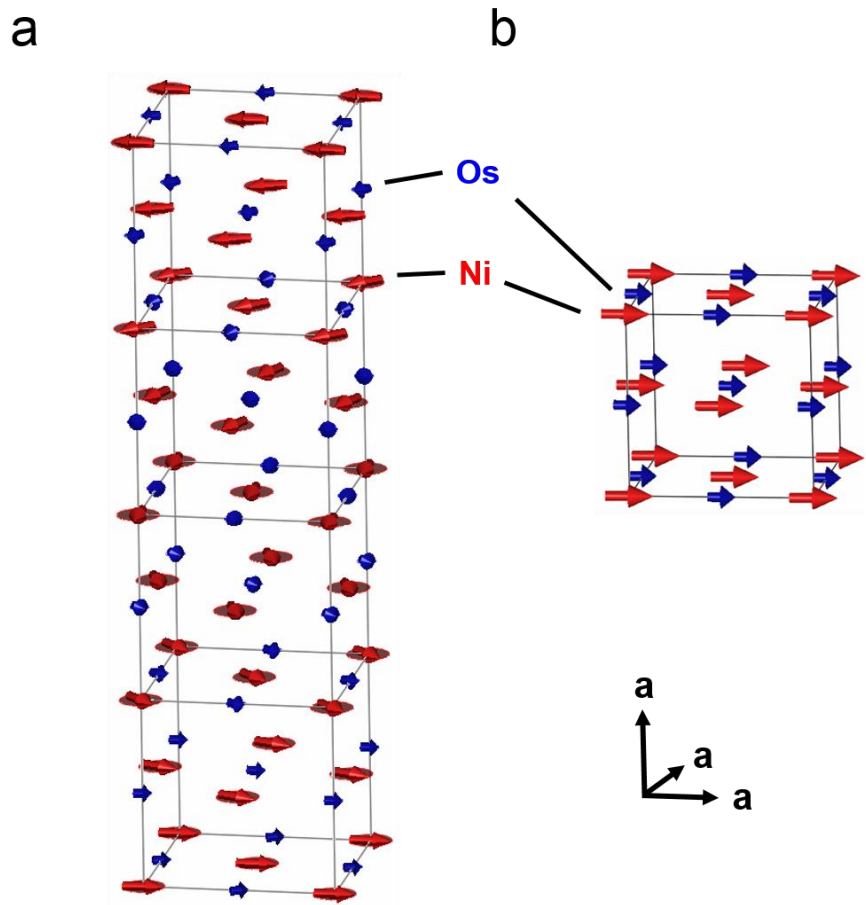

Figure S6. Magnetic order models of (a) the helical and (b) the FM states of $Ba_2NiOsO_6$, depicted from the analysis of the ND profiles. The modulated AFM unit cell consists of 8 nuclear unit cells along one of the cubic lattice axes. The magnetic moments are aligned perpendicular to the propagation vector.

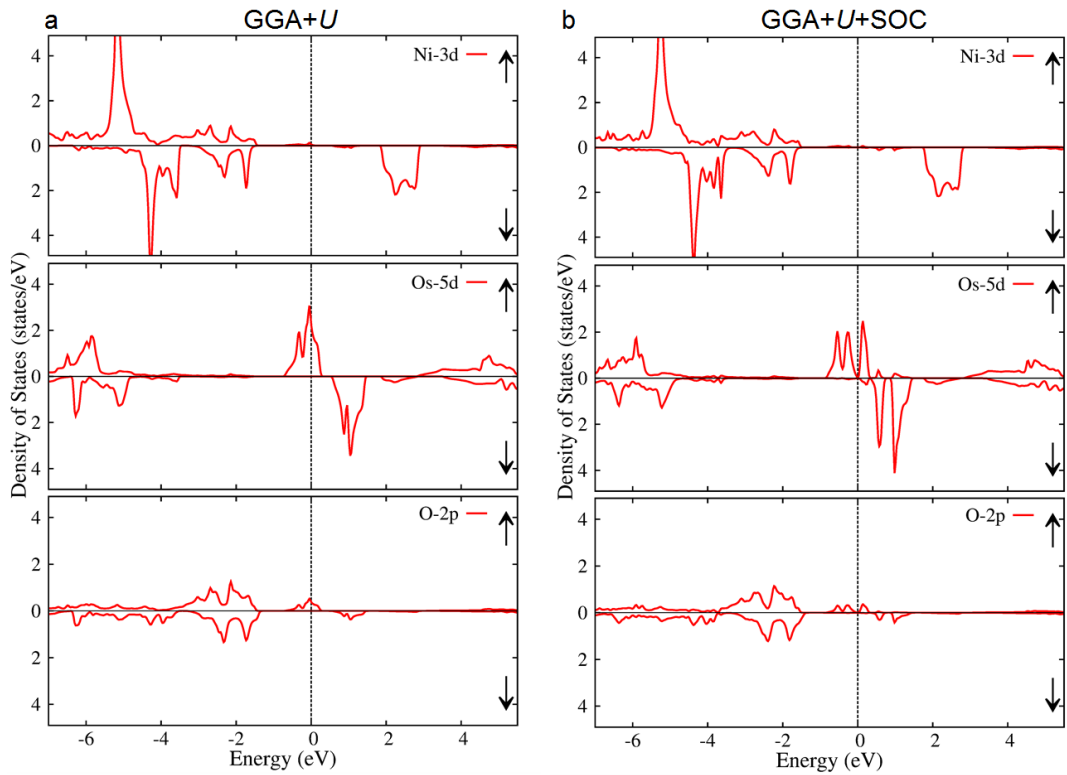

Figure S7. Partial DOS of the Ni-3d, Os-5d, and O-2p bands obtained by (a) GGA+$U$ and (b) GGA+$U$+SOC schemes for spin-up (↑) and spin-down (↓) channels of the FM1 state of Ba$_2$NiOsO$_6$. Vertical dotted line indicates $E_F$ set at zero.

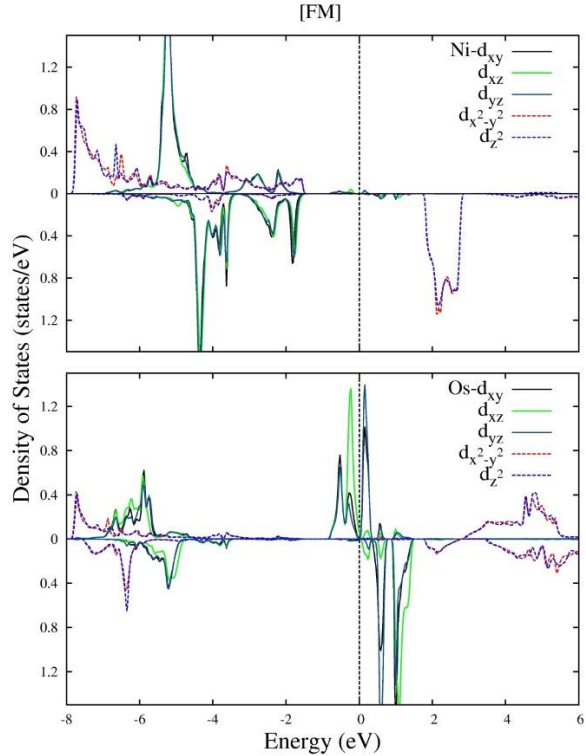

Figure S8. Partial DOS with resolution of $m_l$ contributions of the Ni-3d and Os-5d states within GGA+$U$+SOC, projected to spin-up and spin-down channels of the FM1 state of $Ba_2NiOsO_6$. Vertical dotted line indicates $E_F$ set at zero. Full lines are related to $t_{2g}$ symmetry in the case without SOC, dotted lines to $e_g$.

**Supplemental description of the possible helical magnetic order:** The AF2 order is similar to the experimental helical state (Figure S6a) with ferromagnetic spin arrangement within the $x$–$y$ planes in both cases. The difference between both cases lies in the helical angle of 22.5° (experiment) vs. 180° (model). It is possible that a calculation with a smaller helical angle would yield a lower total energy. We refrained from further calculations, though, considering the relatively large computational effort (up to 16 chemical unit cells) and regarding the AF2 order as a decent model for the experimental ground state.

The spin-resolved total density of states (DOS) for AF2 order is shown in Figure S9 with and without the presence of SOC. Without considering the SOC, $Ba_2NiOsO_6$ is found to be metallic both in GGA and GGA+$U$ for all considered values of $U$ (up to 4 eV for Os). Although the metallic state is robust if SOC is disregarded, a gap opens at $E_F$ if SOC is considered at $U(Os) \geq 1.5$ eV, i.e., by a combined action of SOC and Coulomb on-site interaction of a sufficient strength. The calculated charge gap seems to account for the experimentally observed semiconducting behavior.

We find that a charge gap opens also in the AF2 state, if both SOC and Coulombic correlations are considered simultaneously, see Figure S9. The mechanism of this effect is the same as explained before for the case of FM1, which becomes obvious from the related band dispersions, Figure S10, in conjunction with the partial DOS, Figures S11 and S12. To assist with the understanding of the total DOS structure, simplified partial DOS schemes are depicted in Figure S13. Note that the experimental gap of 0.31 eV is reproduced at a $U$-value of 3 eV in the AF2 state. Summarizing this discussion, the decent agreement between AF2 electronic structure and the experimental results suggests that $Ba_2NiOsO_6$ is possibly a Dirac-Mott insulator (in zero field), where both Coulomb and relativistic effects must be considered equally, similar to the cases of $Sr_2IrO_4$,[S36-38] α-$RuCl_3$,[S39] and $Ca_3CoMO_6$ ($M$ = Co, Rh).[S40-42]

Next, we consider the magnitude of the magnetic moments of the AF2 state. The atomic spin magnetic moments are found to be ±1.78 $\mu_B$ per Ni and ∓1.03 $\mu_B$ per Os, and the related orbital contributions are ±0.17 $\mu_B$ and ±0.44 $\mu_B$, respectively. The relatively larger orbital moment of Os is likely caused by the larger spin-orbit coupling of Os in comparison with Ni. Comparing these values with the experimental data, we find that the calculated total Ni moment of 1.95 $\mu_B$ is considerably larger than the experimental value [1.2(2) $\mu_B$], while the total Os moment of 0.59 $\mu_B$ almost matches the upper limit of the experimental result of 0.3(2) $\mu_B$. Due to hybridization, also oxygen gains small magnetic moments with values in the order of ±0.1 $\mu_B$ per atom, depending on the specific position. Oxygen spin polarization is mainly found on p orbitals perpendicular to Ni–O–Os bonds, as shown in the isosurface plot in Figure S14.

The magnetically easy axis within AF2 is found along the cubic [110] direction. However, the energy is only ~9 meV/f.u. larger, if the moments point along the cubic [100] direction, see Table 2. Thus, we can consider the situation as being consistent with the experimental result that the moments are aligned perpendicular to the propagation vector. The magnetic hard axis is [001], with an energy of 15–25 meV/f.u. higher than the in-plane energy. In the FM state, the calculated easy axis is [110]. The [100] direction of magnetization, which is the experimental easy axis, is found at 13-18 meV/f.u. higher energy in the calculation (see Table 2 and the discussion above).

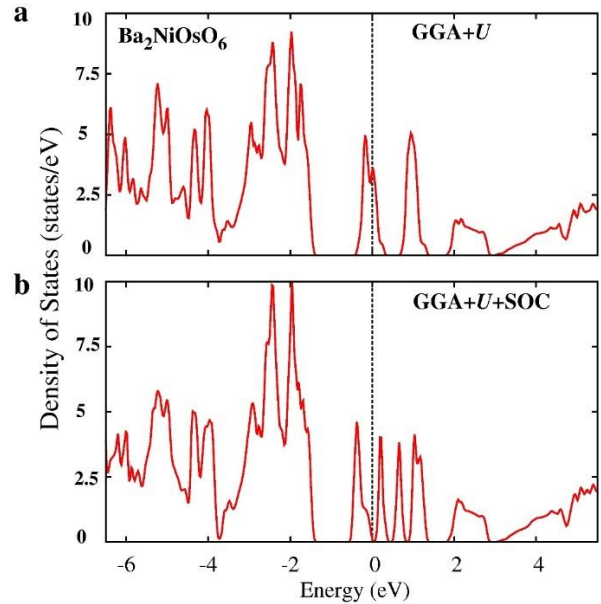

Figure S9. Total DOS (per f.u.) for the AF2 state of $Ba_2NiOsO_6$ obtained within the (a) GGA+$U$ and (b) GGA+$U$+SOC schemes. The vertical dotted line indicates the $E_F$ set at zero. To avoid redundant information, only one spin channel is shown.

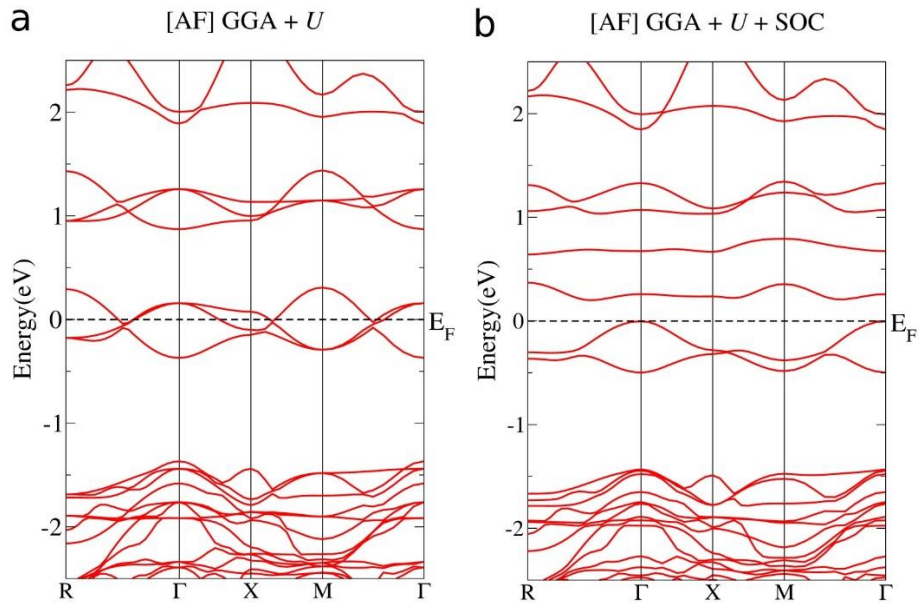

Figure S10. Band dispersion of the AF2 state of $Ba_2NiOsO_6$ in (a) GGA+$U$ and (b) GGA+$U$+SOC schemes.

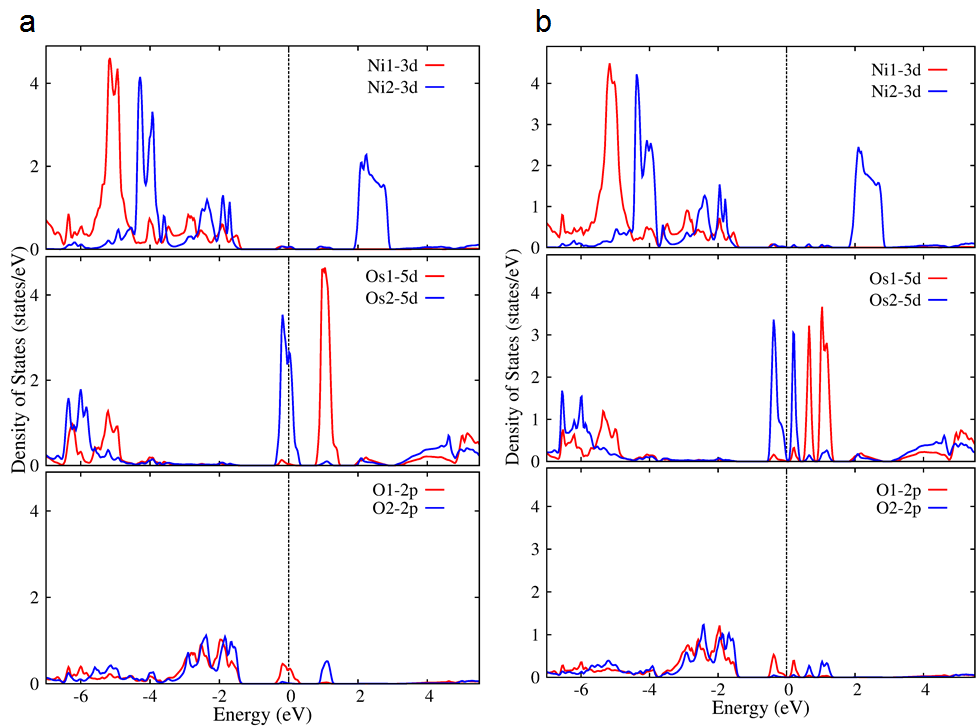

Figure S11. Partial DOS of the Ni-3d, Os-5d, and O-2p bands obtained by (a) GGA+$U$ and (b) GGA+$U$+SOC schemes for the AF2 state of $Ba_2NiOsO_6$. Vertical dotted line indicates $E_F$ set at zero. To avoid redundancy, only one spin channel is shown. The DOS of the other spin is obtained by exchanging the red and blue curves.

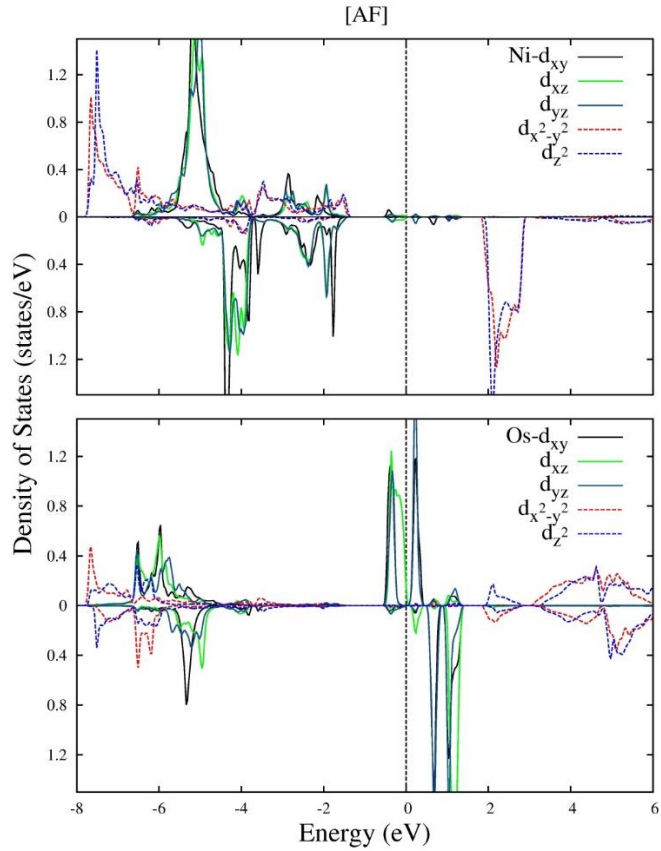

Figure S12. Partial DOS with resolution of $m_l$ contributions of the Ni(1)-3d and Os(2)-5d states within GGA+$U$+SOC, projected to spin-up and spin-down channels of the AF2 state of Ba$_2$NiOsO$_6$. Vertical dotted line indicates $E_F$ set at zero. Full lines are related to $t_{2g}$ symmetry in the case without SOC, dotted lines to $e_g$.

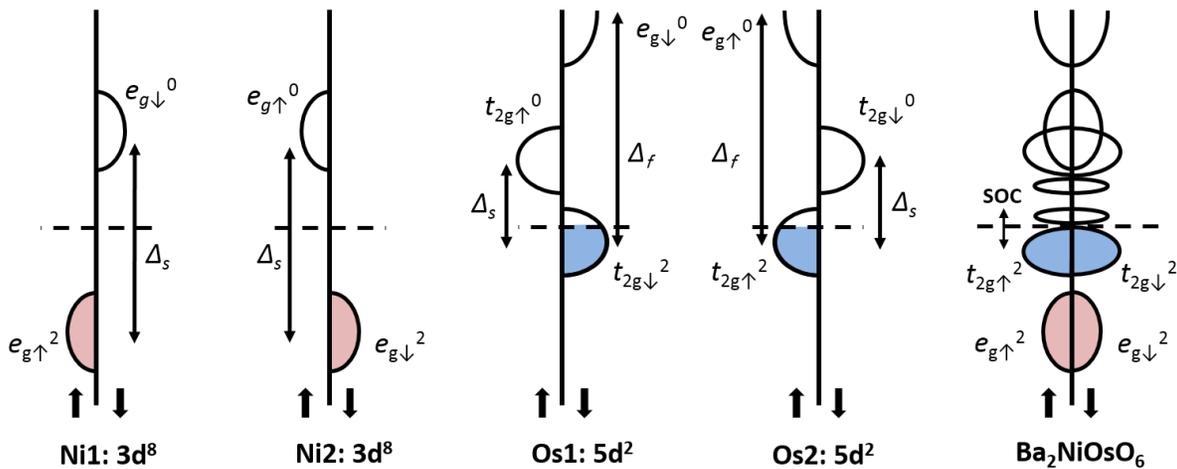

Figure S13. Schematic images of the DOS structures of d bands of the AF2 state of Ba$_2$NiOsO$_6$. The symbols $\Delta_s$ and $\Delta_f$ represent spin exchange splitting and crystal field splitting (including correlation effects), respectively.

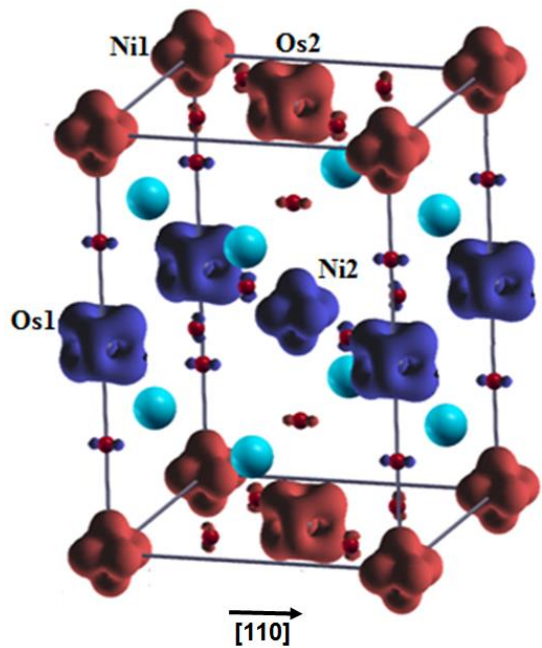

Figure S14.  Isosurface of angular magnetization density at ±0.23 e/Å$^3$ with red (blue) for spin up (down) of the AF2 state of Ba$_2$NiOsO$_6$, calculated within GGA+$U$+SOC and with moments oriented along the cubic [100] direction. In addition, spheres are shown at the atomic positions.